\documentclass{PoS}

\newcommand{\bc}{\begin{center}}
\newcommand{\ec}{\end{center}}
\newcommand{\beqn}{\begin{equation}}
\newcommand{\eeqn}{\end{equation}}
\newcommand{\barr}{\begin{eqnarray}}
\newcommand{\earr}{\end{eqnarray}}

\def\simge{\mathrel{%
       \rlap{\raise 0.511ex \hbox{$>$}}{\lower 0.511ex \hbox{$\sim$}}}}
\def\simle{\mathrel{
       \rlap{\raise 0.511ex \hbox{$<$}}{\lower 0.511ex \hbox{$\sim$}}}}

\def\SR    {\mathord{\buildrel{\lower3pt\hbox{$\scriptscriptstyle\rightarrow$}}\over S}}
\def\SL    {\mathord{\buildrel{\lower3pt\hbox{$\scriptscriptstyle\leftarrow$}}\over S}}

\title{Static quark free energies at finite temperature 
with two flavors of improved Wilson quarks}

\ShortTitle{Static quark free energies at finite temperature 
with two flavors of improved Wilson quarks}

\author{\speaker{Y.~Maezawa}, S.~Ejiri, T.~Hatsuda, N.~Ishii, N.~Ukita\\

        Department of Physics, The University of Tokyo, \\
        Bunkyo-ku, Tokyo 113-0033, Japan

        E-mail: \email{maezawa@nt.phys.s.u-tokyo.ac.jp}}

\author{S.~Aoki and K.~Kanaya \\

        Graduate School of Pure and Applied Sciences, University of Tsukuba, \\
        Tsukuba, Ibaraki 305-8571, Japan
        }

\abstract{
Polyakov loop correlations at finite temperature in two-flavor QCD
are studied in lattice simulations with the RG-improved gluon action
and the clover-improved Wilson  quark action. 
 From the simulations on a  $ 16^3 \times 4$ lattice , we
 extract  the free energies, the  effective running coupling $g_{\rm eff}(T)$  and the Debye
screening mass $m_D(T)$  for  various color channels of heavy
 quark$-$quark
 and quark$-$anti-quark  pairs
 above the critical temperature.  The free energies are well
  approximated by the screened Coulomb form with 
  the appropriate  Casimir factors. 
 The magnitude and the temperature dependence of the 
 Debye mass are compared to those of the  next-to-leading order
  thermal perturbation theory and to a phenomenological
   formula given in terms of  $g_{\rm eff}(T)$.
 Also we made a comparison between  our results with  the Wilson quark and 
 those   with  the staggered quark  previously reported. 
}

\FullConference{XXIVth International Symposium on Lattice Field Theory\\

                July 23-28, 2006\\

                Tucson, Arizona, USA}

\begin{document}

\section{Introduction}

The  interaction between 
heavy quarks in a hot QCD medium 
 is one of the most important quantities to characterize
 the properties of  the quark-gluon plasma (QGP).
 Experimentally, it is related to  the fate of 
 the charmoniums and bottomoniums  
 in the QGP created in relativistic heavy ion collisions.
  In this report, we present our recent studies on
   the  heavy-quark free energy in dynamical simulations
  of two-flavor QCD with the Wilson fermion.
  We study the free energy of a quark ($Q$) and an antiquark ($\overline{Q}$)
  separated by the spatial distance $r$  in the color singlet and octet channels,
   and also study the free energy  of $Q$ and $Q$ 
 in the  color anti-triplet and sextet channels.
 We adopt the  Coulomb gauge fixing  for  the gauge-non-singlet free energies. 
   By fitting the numerical results  with the screened Coulomb form,
 we extract an effective running coupling and 
 the Debye screening mass in each channel as a function
 of temperature.
 
 We find that 
 (i) the free energies in the different channels at high temperature
  ($T \simge 2 T_c$) 
 can be well described by the channel-dependent 
 Casimir factor together with  the channel-independent
  running coupling
  $g_{\rm eff}(T)$ and the Debye mass $m_{\rm D}(T)$,
 (ii)  the next-to-leading order result of the Debye mass in 
  thermal perturbation gives better agreement 
  with $m_D(T)$ on the lattice than that of the leading order,
  (iii) $g_{\rm eff}(T)$ and $m_D(T)$ may be related through the 
 leading order  formula, $m_D(T)= \sqrt{1+N_f/6} \ g_{\rm eff} (T) T$,
  so that most of the higher order and non-perturbative 
   effects on the Debye mass may be   absorbed in $g_{\rm eff}(T)$
  for $T> 1.5T_c$, 
    and
(iv) there is a quantitative discrepancy of $m_D(T)$ 
between our results using  the Wilson quark and those  using 
 the staggered quark even at $T\sim 4 T_c$.

\section{Lattice action and simulation parameters}

 We employ
the renormalization group improved gluon action and 
clover improved Wilson quark action with two-flavors.
 The line of constant physics,
 on which the  quark mass (or the ratio of pseudo-scalar and vector meson masses) 
 is kept fixed
 in the space of the coupling $\beta$ and the hopping parameter $K$,
has been studied in refs.~\cite{AliKhan:2000}
using the same action and is discussed further  in ref.~\cite{Ukita}.
 We perform the simulations on the lines of constant physics
 for the quark masses corresponding to $m_\pi / m_\rho = 0.65$ and 0.80.
 Ten (seven) different  temperatures are taken in the interval  
 $T=1.00T_c \sim 4.02 T_c$ 
 for $m_\pi / m_\rho = 0.65$
 ($T=1.07T_c \sim 3.01 T_c$ for $m_\pi / m_\rho = 0.80$)
on a $N_s \times N_t = 16^3 \times 4$ lattice. 
The hybrid Monte Carlo algorithm is employed 
to generate full QCD configurations, and 
 the free energy in each color channel is
measured using 500 configurations at every ten trajectories
after thermalization of 400-1000 trajectories.
The statistical errors are determined by a jackknife method with 
the bin-size of 10 configurations.

\section{Heavy quark free energies}

The free energy of static quarks on a lattice
may be described by the correlations of the  Polyakov loop:
$\Omega ( {\bf x} )  = \prod_{ \tau = 1}^{N_t} U_4 (\tau, {\bf x})$
 where  $N_t$ is a lattice size
in the temporal direction, and 
the $U_\mu (\tau, {\bf x}) \in SU(3) $ is the link variable.
 By appropriate gauge fixing (such as the 
 Coulomb gauge fixing),
one can define the free energy in various 
 color channels separately
\cite{Nadkarni:1986}:
 the color singlet $Q\overline{Q}$ channel ({\bf 1}),
 the color octet $Q\overline{Q}$ channel ({\bf 8}),
 the color anti-triplet $QQ$ channel (${\bf 3}^*$),
 and the color sextet $QQ$ channel ({\bf 6}) :
\barr
e^{-F_{\bf 1}(r,T)/T}
 &=&
  \frac{1}{3} \langle {\rm Tr} 
\Omega^\dagger({\bf x}) \Omega ({\bf y})
\rangle
, 
\label{eq:singlet}
\\
e^{-F_{\bf 8}(r,T)/T}
 &=& 
\frac{1}{8} \langle {\rm Tr} \Omega^\dagger({\bf x})
{\rm Tr} \Omega ({\bf y})
\rangle
-
\frac{1}{24} \langle {\rm Tr} \Omega^\dagger({\bf x})
 \Omega ({\bf y}) \rangle
, \\
e^{-F_{\bf 6}(r,T)/T}
 &=& 
\frac{1}{12} \langle {\rm Tr} \Omega({\bf x})
{\rm Tr} \Omega ({\bf y})
\rangle
+
\frac{1}{12} \langle {\rm Tr} \Omega({\bf x})
 \Omega ({\bf y}) \rangle
, \\
e^{-F_{{\bf 3}^*}(r,T)/T}
 &=& 
\frac{1}{6} \langle {\rm Tr} \Omega({\bf x})
{\rm Tr} \Omega ({\bf y})
\rangle
-
\frac{1}{6} \langle {\rm Tr} \Omega({\bf x})
 \Omega ({\bf y}) \rangle
,
\label{eq:sextet}
\earr
where $r = |{\bf x} - {\bf y}|$.

We introduce  normalized free energies 
$(V_{\bf 1}, V_{\bf 8}, V_{\bf 6}, V_{{\bf 3}^*})$
which are expected to approach zero at large distances 
above $T_c$. 
 This is equivalent to defining the free energies by 
 dividing the right-hand sides of 
Eq.~(\ref{eq:singlet}) -- (\ref{eq:sextet})
by $\langle {\rm Tr} \Omega \rangle^2$.

\begin{figure}[t]
  \begin{center}
    \begin{tabular}{cc}
    \includegraphics[width=73mm]{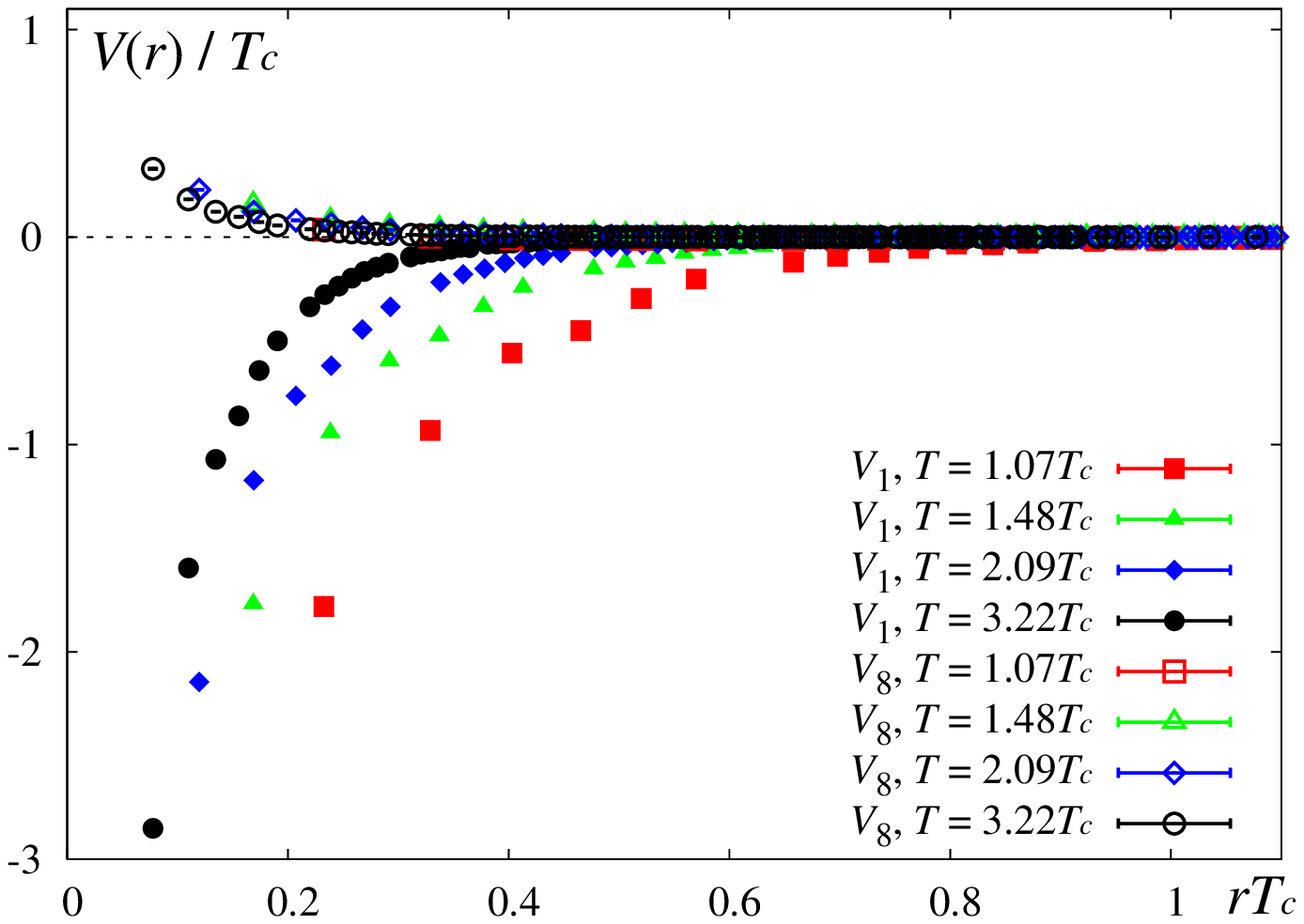} &
    \includegraphics[width=73mm]{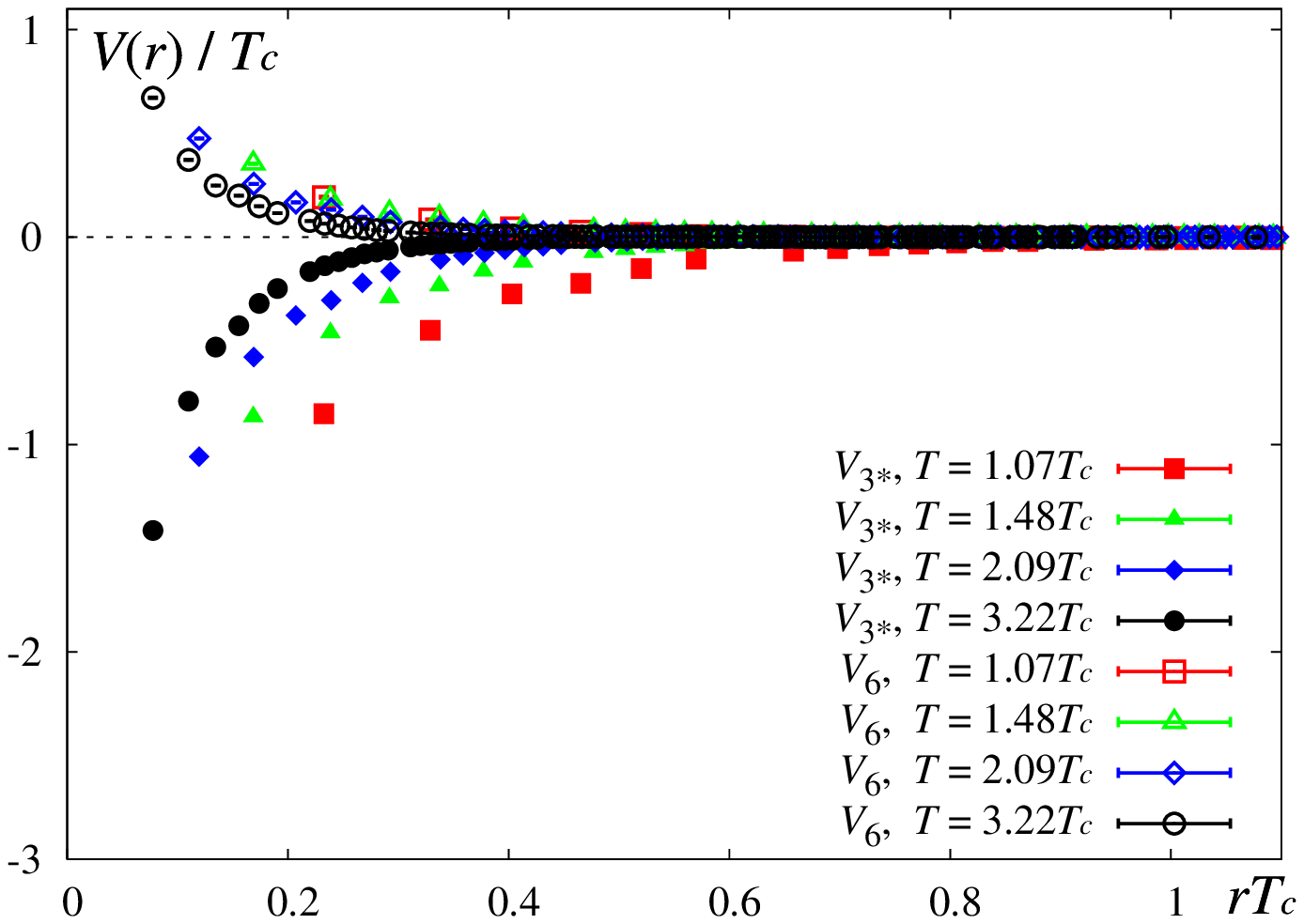}
    \end{tabular}
    \caption{Simulation results of the 
    normalized free energies scaled by $T_c$
    for color singlet and octet $Q\overline{Q}$ channels (left) and 
    color anti-triplet and sextet $QQ$ channels (right)
    at $m_\pi / m_\rho = 0.65$ and several temperatures.
        }
    \label{fig:pot}
  \end{center}
\vspace{-3mm}
\end{figure}

The normalized free energies are shown in Fig.~\ref{fig:pot}
for color singlet and octet $Q\overline{Q}$ channels (left) and 
color anti-triplet and sextet $QQ$ channels (right)
 for  $m_\pi / m_\rho = 0.65$ and $T \ge T_c$.
They are ``attractive''  in the color singlet and anti-triplet channels
and ``repulsive''  in the color octet and sextet channels.
Also  they become weak at long distances as $T$ increases
 due to the effect of  Debye screening.
These behaviors are qualitatively similar to the case in the 
 quenched simulations with the 
 Lorenz gauge reported in ref.~\cite{Nakamura:2004}.

To study the screening effects in  each color channel more closely,
we fit the free energies by the screened Coulomb form:
\begin{equation}
V_M(r,T) = C(M) \frac{\alpha_{\rm eff}(T)}{r} e^{-m_D(T) r} ,
\label{eq:SCP}
\end{equation}
where $\alpha_{\rm eff}(T)$ and $m_D(T)$ 
are the effective running coupling and 
 Debye screening mass, respectively.
The $C(M) \equiv \langle \sum_{a=1}^{8} t_1^a\cdot t_2^a \rangle_M$
 is the Casimir factor for each color channel ($M$) 
defined as  
\barr
C({\bf 1})     = -\frac{4}{3}, \ \ \
C({\bf 3}^*)   = -\frac{2}{3}, \ \ \
C({\bf 8})     =  \frac{1}{6}, \ \ \
C({\bf 6})     =  \frac{1}{3}.
\earr
Here, it is worth stressing that, with the improved actions we adopt,
the rotational symmetry is well restored in the heavy quark free
energies \cite{Aoki:1999ff}.
 Therefore
 we do not need to introduce 
 terms correcting lattice artifacts at short distances 
 in eq.~(\ref{eq:SCP}) to
fit the data shown in Fig.1.

 The Debye screening effect is defined through
the long distance behavior 
of $V_M(r,T)$.
In order to determine  the appropriate fit range,  
we estimate the effective Debye  mass
from a ratio of normalized free energies:
\barr
 m_D(T;r) = \frac{1}{\Delta r} \log \frac{V_M(r)}{V_M(r+ \Delta r)}
 - \frac{1}{\Delta r} \log \left[ 1 + \frac{\Delta r}{r} \right]
.
\earr
 Investigating the plateau of $m_D(T;r)$,
we choose the fit range 
to be $ \sqrt{11}/4 \le rT \le 1.5$.
 Systematic errors due to the difference of the fit range
are about 10\%
for $T \simge 2T_c$.

\begin{figure}[t]
  \begin{center}
    \begin{tabular}{cc}
      \resizebox{73mm}{!}{\includegraphics{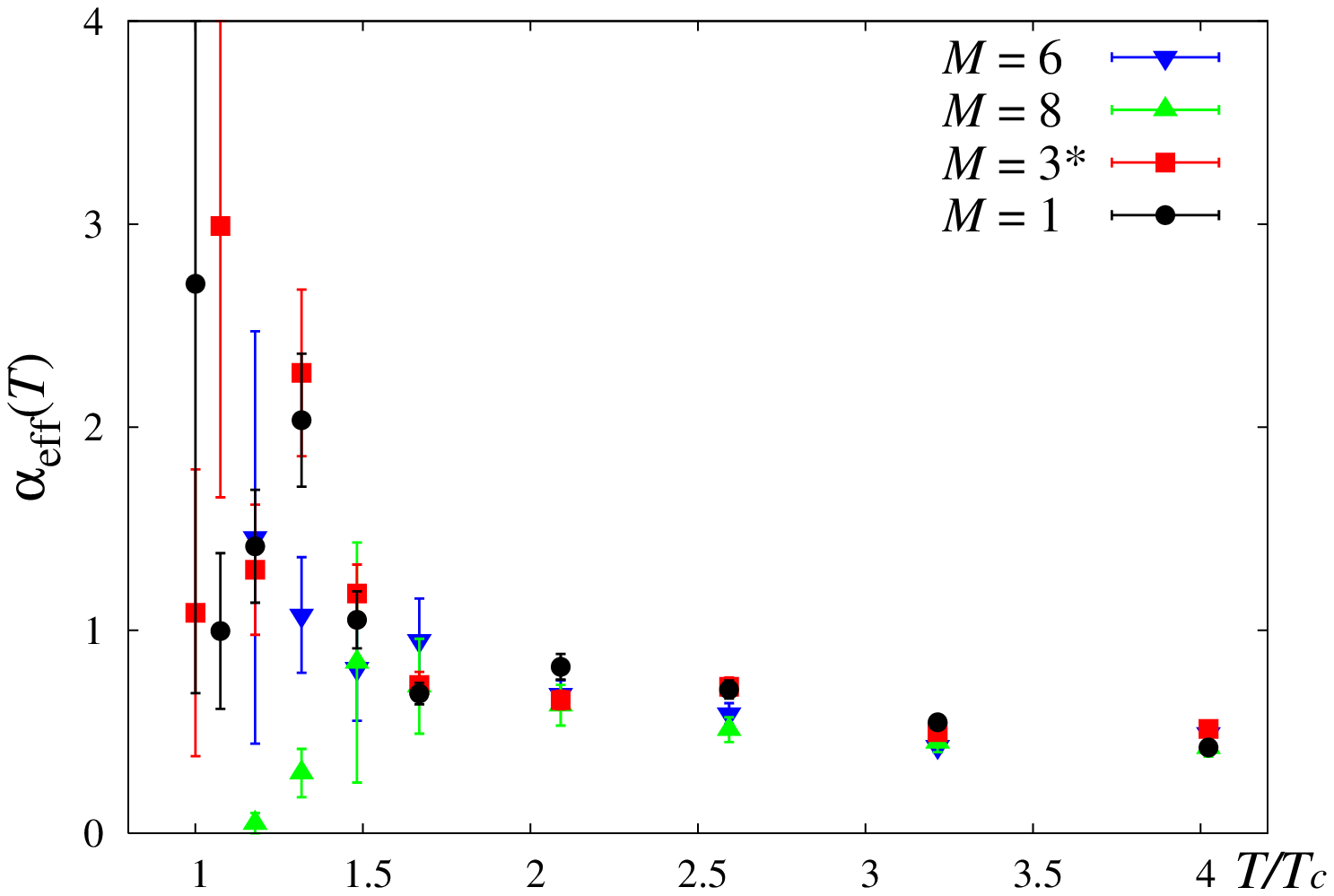}} &
      \resizebox{73mm}{!}{\includegraphics{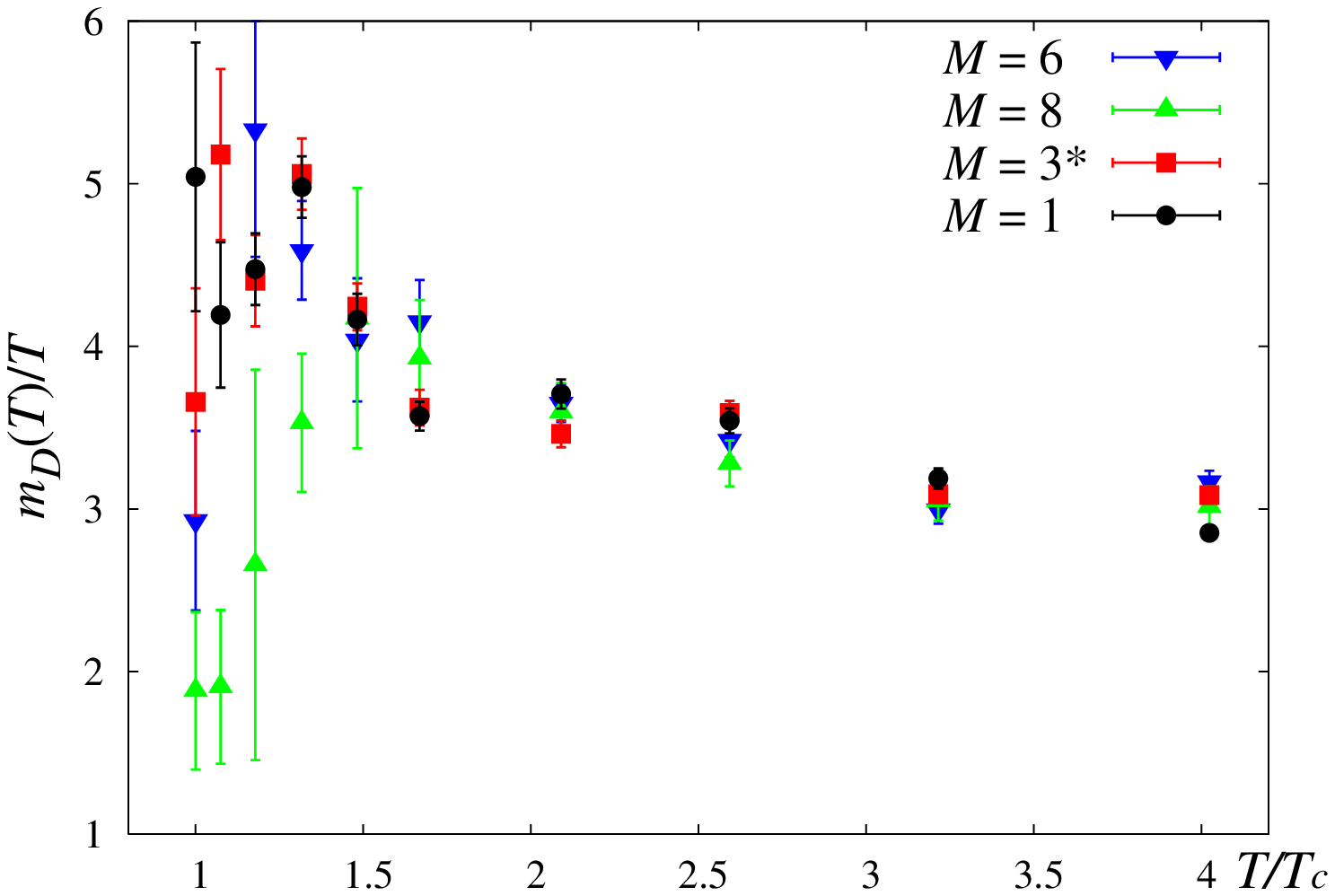}} 
    \end{tabular}
    \caption{The effective running coupling $\alpha_{\rm eff}(T)$ (left) and
    Debye screening mass $m_D(T)$ (right) for each color channel
    as a function of temperature
    from the large distance behavior of the potentials
    at $m_\pi / m_\rho = 0.65$.
    }
    \label{fig:para065}
  \end{center}
\vspace{-3mm}
\end{figure}

The results of the $\alpha_{\rm eff}(T)$ and $m_D(T)$ are shown
in Fig.~\ref{fig:para065} for $m_\pi / m_\rho = 0.65$.
Similar behavior in both results are obtained
for $m_\pi / m_\rho = 0.80$.
We find that there is no significant  channel dependence of 
 $\alpha_{\rm eff}(T)$ and $m_D(T)$
at sufficiently high temperature $(T \simge 2T_c)$.
In other words, the channel dependence of the free energy
at high temperature may be well absorbed in the kinematical Casimir
 factor as first indicated in  
 a quenched study \cite{Nakamura:2004}.

\section{Debye mass on the lattice and that in perturbative theory}    

 Let us first compare the Debye mass on the lattice
  with that calculated in thermal    perturbation theory.
First of all, the  2-loop running coupling is given by
\barr
g^{-2}_{\rm 2l} (\mu) 
=  \beta_0 \ln \left( \frac{\mu}{\Lambda_{\overline{MS}}} \right)^2 + 
\frac{\beta_1}{\beta_0} 
\ln \ln \left( \frac{\mu}{\Lambda_{\overline{MS}}} \right)^2
,
\earr
where  the argument  in the logarithms
may be written as 
$ \mu / \Lambda_{\overline{MS}} = (\mu/T)(T/T_c)(T_c/\Lambda_{\overline{MS}})$
  where 
  we adopt 
$T_c / \Lambda_{\overline{MS}}^{n_f=2} \simeq 0.656$
for the last factor
\cite{AliKhan:2000,Gockeler:2005rv}.
The renormalization point $\mu$ is 
  assume to be in a range $\mu = \pi T  - 3 \pi T$.
 By using $g_{\rm 2l}$ as a function of $T/T_c$,  
the Debye screening mass  in the leading-order (LO) thermal 
perturbation is given as
$ {m_D^{\rm LO}(T)}/{T} = \sqrt{ 1 + N_f / 6 } \  g_{\rm 2l} (T)$, 
where the effect of the quark mass is neglected.

\begin{figure}[t]
  \begin{center}
    \begin{tabular}{cc}
      \resizebox{73mm}{!}{\includegraphics{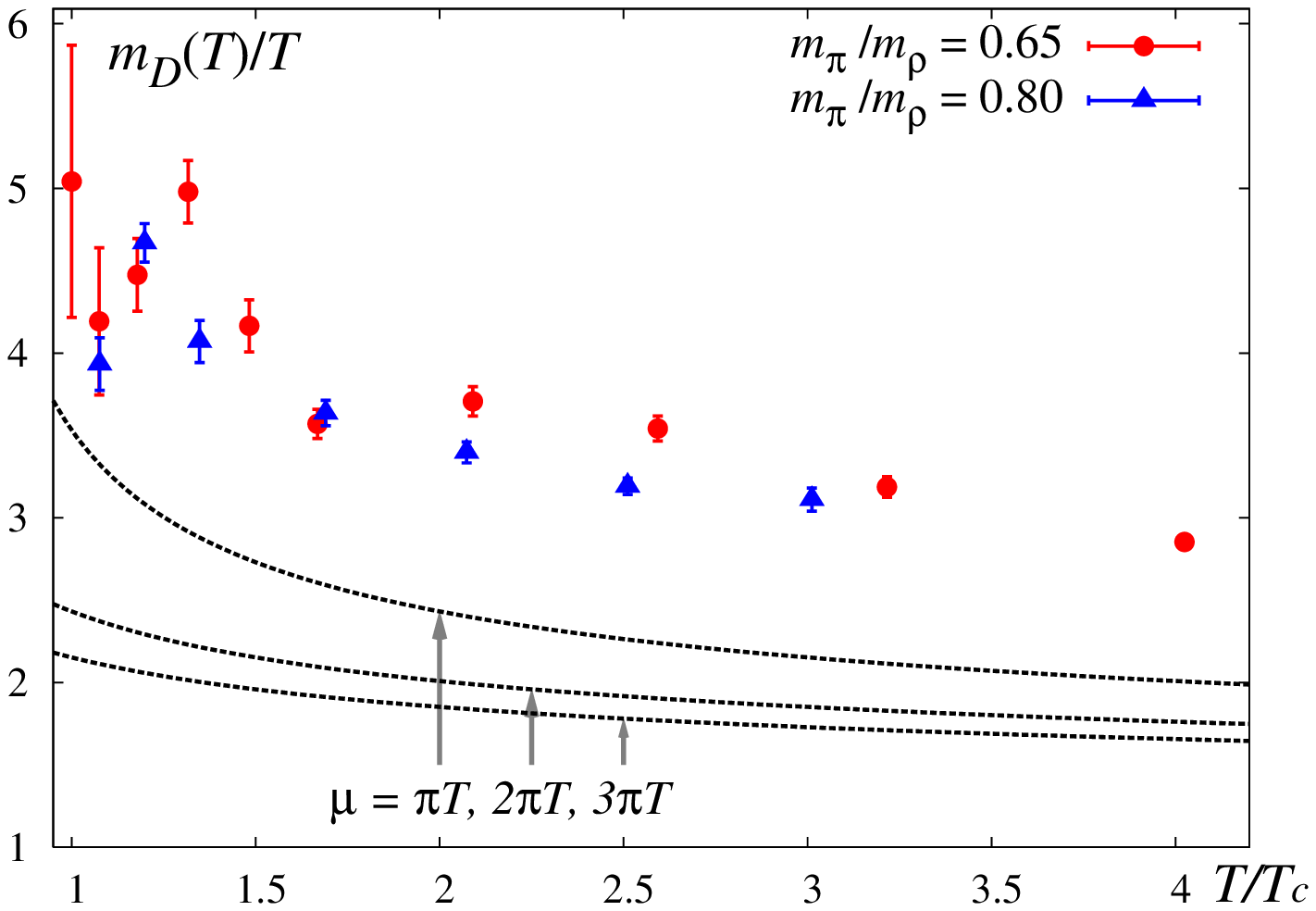}} &
      \resizebox{73mm}{!}{\includegraphics{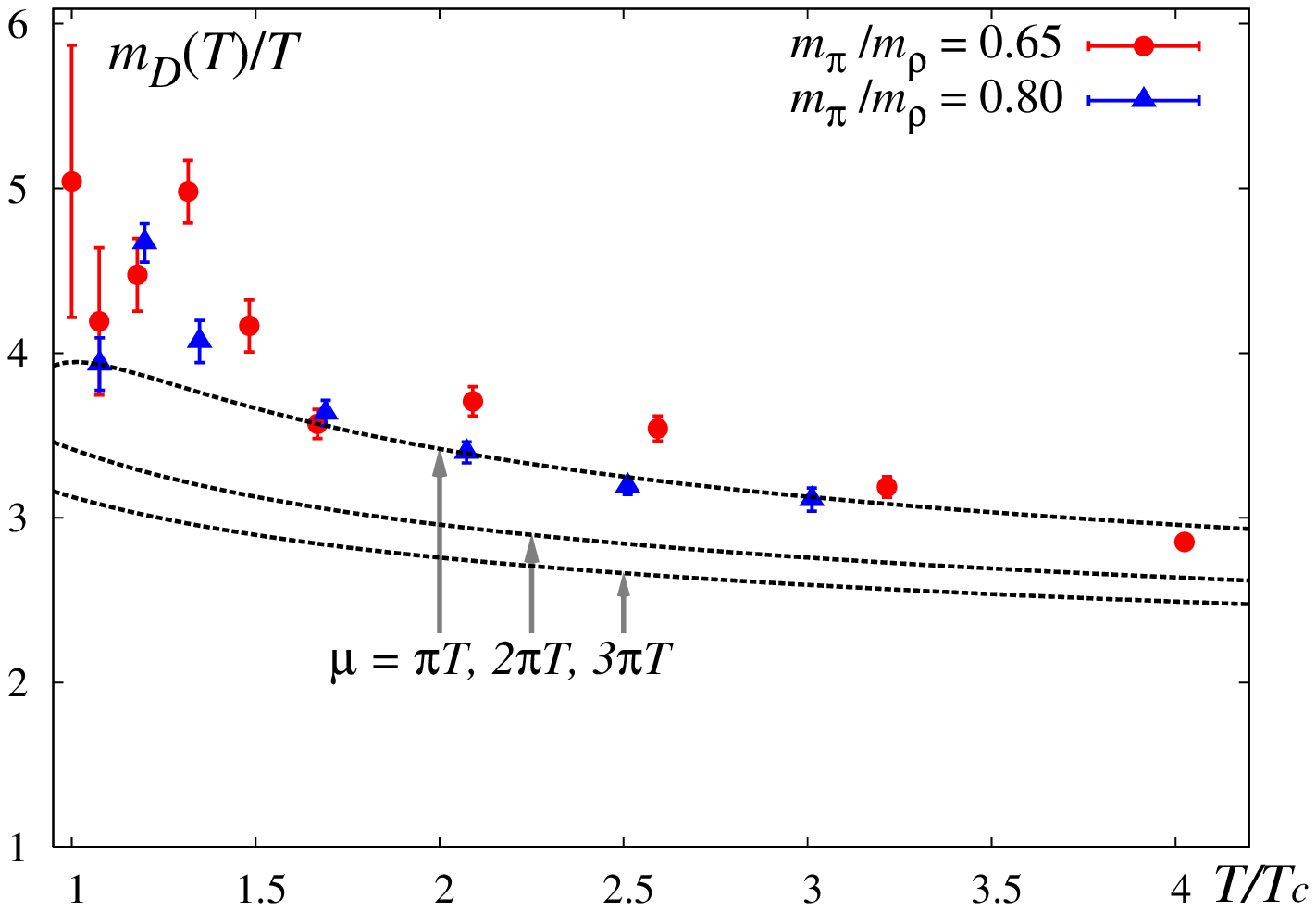}} 
    \end{tabular}
    \caption{The Debye screening masses $m_D(T)$ 
    at $m_\pi / m_\rho = 0.65$ and 0.80 in the color singlet channel
    together with that calculated in the leading-order (left)
    and next-to-leading-order (right) thermal perturbation theory 
    shown by the dashed lines. $\mu$ is the renormalization point
    chosen at $\mu = \pi T, \ 2\pi T, \ 3\pi T$.
    }
    \label{fig:pQCD}
  \end{center}
\vspace{-3mm}
\end{figure}

Figure \ref{fig:pQCD}(left) shows the $m_D(T)$ 
 in the color singlet channel
compared with ${m_D^{\rm LO}(T)}/{T}$
for $\mu = \pi T, \ 2\pi T$ and $3\pi T$.
We find that the screening mass 
$m_D^{\rm LO}(T)$
in the leading order perturbation theory
  does not reproduce  the  lattice data, which
  has been known in the quenched QCD  \cite{Nakamura:2003pu}
   and in the 
  full QCD with the staggered quark \cite{Kaczmarek:2005ui} .
 
To study higher-order contributions in 
the thermal perturbation theory,
we consider the Debye mass  in the  next-to-leading-order 
calculated by the hard thermal resummation 
given in ref.~\cite{Rebhan:1993az}.
\barr
\frac{m_D^{\rm NLO}}{T} = 
\sqrt{ 1 + \frac{N_f}{6} } \
g_{\rm 2l}(T) \left[ 1 + 
g_{\rm 2l}(T) \frac{3}{2 \pi} \sqrt{\frac{1}{1+ N_f/6}}
\left(
\ln \frac{2 m_D^{\rm LO}}{m_{\rm mag}} - \frac{1}{2}
\right)
+ o(g^2)
\right]
.
\label{eq:m_D_NLO}
\earr
 Here $m_{\rm mag}$  denotes the magnetic screening mass
assumed to be  of the form $m_{\rm mag}(T) = C_m g^2(T) T$.
Since the factor $C_m$ cannot be determined 
in the perturbation theory due to the  
infrared problem,
we adopt $C_m \simeq 0.482$ calculated in 
quenched lattice simulations \cite{Nakamura:2003pu}
 as a typical value.  ( If we fit $C_m$ from our lattice data 
 $m_D(T=4.02T_c)$ with $\mu=2\pi T$, we obtain $C_m \simeq 0.40$).
  $m_D^{\rm NLO}(T)$ for different choice of the 
   renormalization point  
 is shown in Fig.~\ref{fig:pQCD}(right)
by the  dashed lines  together with the lattice data.
 They have approximately
 50 \%
  enhancement from the leading order results
 and  lead to a better agreement with the lattice data.

\section{Phenomenological relation between $\alpha_{\rm eff}$ and $m_D$}

So far, we have fitted the free energies on the lattice with 
 $\alpha_{\rm eff}$ and $m_D$ as independent parameters.
 Here let us introduce an ``effective'' running coupling
   as $g_{\rm eff} (T) \equiv \sqrt{4 \pi \alpha_{\rm eff}(T) } $.
  Suppose that $m_D(T)$ is expressed 
by $g_{\rm eff}(T)$ according to the leading-order perturbation,
\barr
\frac{m_D(T)}{T} = \sqrt{1 + \frac{N_f}{6} } \ g_{\rm eff} (T) .
\label{eq:m_eff}
\earr
This relation means that the following ratio $R$  evaluated 
from our lattice data should be close to unity, i.e.
$R(T) \equiv   ( 1 + N_f / 6 )^{-1/2} \ (m_D(T)/T) /
{ \sqrt{4 \pi \alpha_{\rm eff}(T)} }  \sim 1$.

\begin{figure}[t]
  \begin{center}
      \resizebox{73mm}{!}{\includegraphics{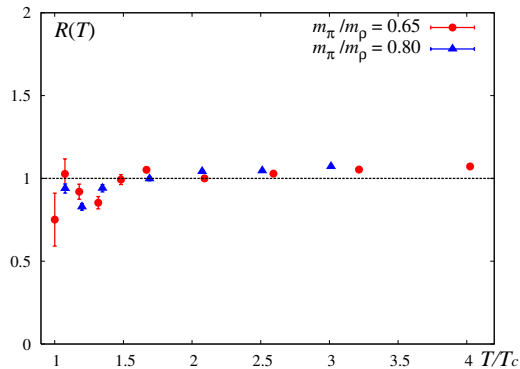}} 
     \caption{
     The ratio $R(T)$ in the text which is 
      supposed to be close to unity if  
      $m_D(T) = \sqrt{1 + N_f/6 } \ g_{\rm eff} (T) T$ 
      holds.
     The data 
        are for the color singlet channel. 
        }
     \label{fig:P_ratio}
 \end{center}
\vspace{-3mm}
\end{figure}

In Fig.~\ref{fig:P_ratio},  the lattice data of $R(T)$ 
for the singlet channel are shown as a function of $T$.
We find that $R(T)$ is consistent  with unity 
 even at $ T \simge 1.5 T_c$ with 10\% accuracy.
This is a non-trivial observation particularly near $T_c$
 and  suggests that the major part of the higher-order effects
  and non-perturbative effects of $m_D(T)$ can be 
   expressed by the effective running coupling $g_{\rm eff}(T)$. 
We note that a similar effective coupling defined through the lattice
potential was discussed to improve the lattice perturbation theory at
$T = 0$ \cite{Lepage:1992xa}.

\section{Comparison with the staggered quark}

\begin{figure}[t]
  \begin{center}
    \begin{tabular}{cc}
      \resizebox{73mm}{!}{\includegraphics{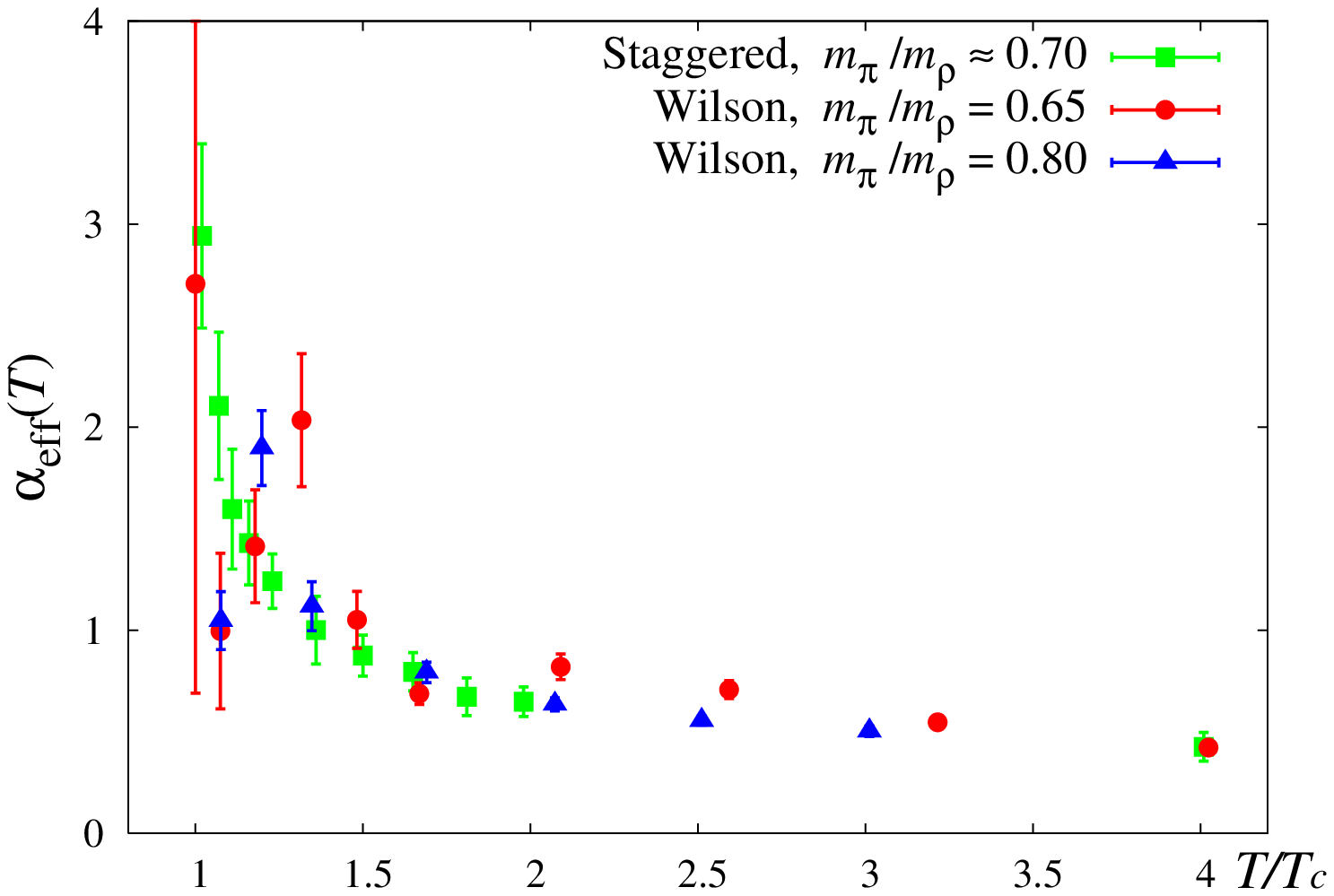}} &
      \resizebox{73mm}{!}{\includegraphics{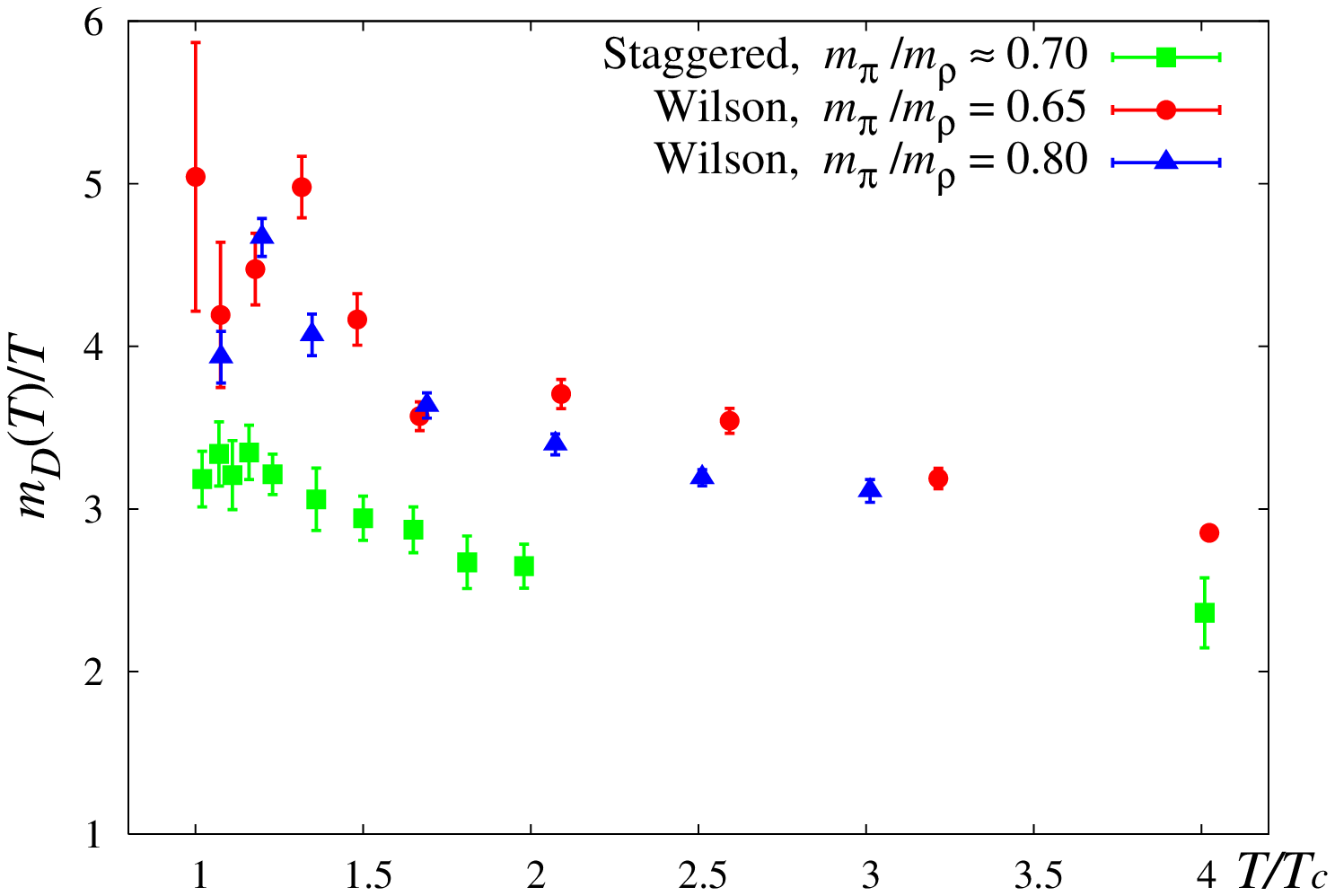}} 
    \end{tabular}
    \caption{Comparison of the $\alpha_{\rm eff}(T)$(left) 
    and $m_D(T)$(right) between
    the results of the Wilson quark action and 
    staggered quark action.
    }
    \label{fig:KS}
  \end{center}
\vspace{-3mm}
\end{figure}

Finally, we compare the results of  $\alpha_{\rm eff}(T)$ and $m_D(T)$
obtained with the Wilson quark action (present work)
with those with an improved staggered quark action.
The latter simulation was done
on a  $16^3 \times 4$ 
and with a quark mass corresponding 
to $m_\pi / m_\rho \simeq 0.70$ \cite{Kaczmarek:2005ui}.
The comparison is shown in
 Fig.~\ref{fig:KS} for  $\alpha_{\rm eff}(T)$ (left panel)
and $m_D(T)$(right panel).
Although  $\alpha_{\rm eff}(T)$ does not show significant
 difference between
the two actions, $m_D(T)$ in the Wilson action is
systematically higher than that of the staggered action 
by about 20\% 
even at $4 T_c$.
 This difference should be further investigated by 
 increasing the temporal lattice size.

\section{Summary}

We have studied the free energy of $QQ$ and $Q \overline{Q}$ systems
 in 2-flavor QCD at finite temperature using the lattice simulation with
the renormalization group improved gluon action and the  
clover improved Wilson quark action on a $16^3 \times 4$ lattice.
The free energy normalized to be zero
 at large separation show attraction (repulsion)
  in the color singlet and anti-triplet channels (color octet and sextet
 channels).

The screened Coulomb form with the Casimir factor 
 and with the effective coupling $\alpha_{\rm eff}(T)$
  and the Debye screening mass $m_D(T)$ as free parameters 
 is used to fit the free energy in each channel.
 $\alpha_{\rm eff}(T)$ and $m_D(T)$ 
 become universal and 
all the channel dependence is absorbed in the 
 Casimir factor for $T \simge 2 T_c$. 
The magnitude and the $T$-dependence of the
  Debye mass $m_D(T)$ is consistent
 with the next-to-leading order calculation in the
 perturbation theory
 as well as the leading order perturbation 
 with the ``effective'' running coupling.

The results from an improved Wilson quark action
are compared with these from an improved 
staggered quark action
with the same lattice size and similar quark mass.
The $\alpha_{\rm eff}(T)$ does not show appreciable difference
between the two actions, whereas
the $m_D(T)$ of the Wilson quark action is systematically
higher than that of the staggered quark action by 20\%.
 The simulations with 
 larger lattice sizes 
especially in the temporal direction (such as $N_t=6$ and larger)
should be carried out 
as well as those at small quark masses.

\paragraph{Acknowledgements:}
We thank O.~Kaczmarek for providing us the data
from a staggered quark action.
This work is in part supported 
by Grants-in-Aid of the Japanese Ministry
of Education, Culture, Sports, Science and Technology, 
(Nos.~13135204, 15540251, 17340066,
18540253, 18740134). SE is supported by the Sumitomo Foundation (No.~050408),
 and YM is supported
by JSPS. This work is in part supported 
also by the Large-Scale Numerical Simulation
Projects of ACCC, Univ.~of Tsukuba.


\end{document}